\documentclass[preprint,12pt]{elsarticle}
\usepackage{amssymb}
\usepackage{comment}
\usepackage[T1]{fontenc}
\usepackage{graphicx}
\usepackage{tikz}
\usetikzlibrary{arrows}

\usepackage{fullpage}
\usepackage{lineno}

\usepackage{xspace}
\usepackage{color}
\usepackage{amsmath}
 
\usepackage{balance}
\usepackage{algorithm}
\usepackage[noend]{algorithmic} 
\usepackage{multirow}
\usepackage{mathtools}
\DeclareFontFamily{U}{mathb}{\hyphenchar\font45}
\DeclareFontShape{U}{mathb}{m}{n}{
      <5> <6> <7> <8> <9> <10> gen * mathb
      <10.95> mathb10 <12> <14.4> <17.28> <20.74> <24.88> mathb12
      }{}
\DeclareSymbolFont{mathb}{U}{mathb}{m}{n}
\DeclareMathSymbol{\dotdiv}{2}{mathb}{"01}

\usepackage{caption} 
\usepackage{subcaption}
\usepackage{amsthm}
\theoremstyle{plain}
\newtheorem{theorem}{Theorem}
\newtheorem{lemma}{Lemma}

\newcommand{\Wait}{{\tt Wait}\xspace}
\newcommand{\Look}{{\tt Look}\xspace}
\newcommand{\Compute}{{\tt Compute}\xspace}
\newcommand{\Move}{{\tt Move}\xspace}
\newcommand{\LCM}{{\tt LCM}\xspace}

\newcommand{\pre}{\mathtt{pre}}
\newcommand{\fsync}{\textsc{FSync}\xspace}
\newcommand{\ssync}{\textsc{SSync}\xspace}

\newcommand{\async}{\textsc{Async}\xspace}

\newcommand{\Aut}{\mathrm{Aut}}

\newcommand{\Ex}{\mathbb{E}} 
\newcommand{\A}{\mathcal{A}} 

\newcommand{\tc}{\mathit{tc}}
\newcommand{\h}{\mathit{H}}

\newcommand{\GS}{G_{\infty}} 
\newcommand{\GSmn}{G}        

\newcommand{\GMV}{\mathrm{GMV}}  
\newcommand{\GMVopt}{\mathrm{GMV\!}_{\mathsf{area}}} 

\newcommand{\algoinfty}{\mathcal{A}_{\mathsf{\infty}}} 
\newcommand{\algo}{\mathcal{A}} 
\newcommand{\algoAsym}{\mathcal{A}_{\mathsf{asym}}} 
\newcommand{\mbr}{\mathit{mbr}}
\newcommand{\LSS}{\mathit{LSS}}
\newcommand{\RS}{\mathit{RS}}
\newcommand{\RV}{\mathit{RC}}
\newcommand{\Bl}{\mathit{B_l}}
\newcommand{\Al}{\mathit{A_l}}
\newcommand{\FO}{\mathit{p}}

\newcommand{\Prop}{ \mathsf{Prop} }
\newcommand{\specialpath}{special-path\xspace}
\newcommand{\specialpaths}{special-paths\xspace}
\newcommand{\specialsubpath}{special-subpath\xspace}
\newcommand{\specialsubpaths}{special-subpaths\xspace}
\newcommand{\false}{\mathtt{false}}
\newcommand{\true}{\mathtt{true}}

\newcommand{\nil}{\mathit{nil}}
\newcommand{\Prob}{\mathcal{P}}

\newcommand{\problemnr}[4]{%
  \begin{problem}
  \begin{small}
  \label{#2}
  {#1}
  \\[-2ex]
  \begin{list}{\ }{%
      \setlength{\topsep}{0ex}
      \setlength{\labelwidth}{4.5em}
      \setlength{\leftmargin}{5em}
      \setlength{\rightmargin}{0em}
      \setlength{\labelsep}{0.5em}
      \setlength{\itemsep}{0ex}
      }
      \item[\textit{Input:}\hfill] #3
      \item[\textit{Goal:}\hfill] #4 \medskip
  \end{list} 
  \end{small}
  \end{problem}
}

\newcommand{\block}[1]{~\\ \noindent {\textsc{#1.}}~}
\definecolor{linecomment}{rgb}{0.95, 0.1, 0.1}
\definecolor{comment}{rgb}{0.1, 0.1, 0.95}

\newenvironment{problem}{%
  \smallskip\noindent\textbf{Problem:}
}{%
}

\definecolor{linecomment}{rgb}{0.95, 0.1, 0.1}
\definecolor{comment}{rgb}{0.1, 0.1, 0.95}

\journal{}

\begin{document}

\begin{frontmatter}


\title{Time-optimal geodesic mutual visibility of robots \\ on grids within minimum area}


\author[inst1]{Serafino Cicerone}
\ead{serafino.cicerone@univaq.it}
\affiliation[inst1]{organization={Dipartimento di Ingegneria e Scienze dell'Informazione e Matematica, Università degli Studi dell'Aquila},
            addressline={\\ Via Veotio}, 
            postcode={I-67100}, 
            city={L'Aquila},
            country={Italy}}

\author[inst1]{Alessia Di Fonso}
\ead{alessia.difonso@univaq.it}
\author[inst1]{Gabriele Di Stefano}
\ead{gabriele.distefano@univaq.it}
\author[inst2]{Alfredo Navarra}
\ead{alfredo.navarra@unipg.it}

\affiliation[inst2]{organization={Dipartimento di Matematica e Informatica, Università degli Studi di Perugia},
            addressline={\\ Via Vanvitelli 1}, 
            postcode={I-06123}, 
            city={Perugia},
            country={Italy}}

\begin{abstract}
The \textsc{Mutual Visibility} is a well-known problem in the context of mobile robots. For a set of $n$ robots disposed in the Euclidean plane, it asks for moving the robots without collisions so as to achieve a placement ensuring that no three robots are collinear. For robots moving on graphs, we consider the \textsc{Geodesic Mutual Visibility} ($\GMV$) problem. Robots move along the edges of the graph, without collisions, so as to occupy some vertices that guarantee they become pairwise geodesic mutually visible. This means that there is a shortest path (i.e., a ``geodesic'') between each pair of robots along which no other robots reside.
We study this problem in the context of finite and infinite square grids, for robots operating under the standard Look-Compute-Move model. 
In both scenarios, we provide resolution algorithms along with formal correctness proofs, highlighting the most relevant peculiarities arising within the different contexts, while optimizing the time complexity.
\end{abstract}



\begin{keyword}
Autonomous mobile robots \sep Oblivious robots \sep Mutual visibility \sep Grids 
\end{keyword}

\end{frontmatter}

\section{Introduction}\label{sec:intro}


We consider swarm robotics concerning autonomous, identical and homogeneous robots operating in cyclic operations. 
Robots are equipped with sensors and motion actuators and operate in standard \emph{Look-Compute-Move} cycles (see, e.g.,~\cite{CDN21a,FPS-macbook19,DDKN12}). When activated, in one cycle a robot takes a snapshot of the current global configuration (\Look) in terms of relative robots' positions, according to its own local coordinate system. Successively, in the \Compute phase, it decides whether to move toward a specific direction or not and in the positive case it moves (\Move). A Look-Compute-Move cycle forms a computational cycle of a robot.
What is computable by such entities has been the object of extensive research within distributed computing, see, e.g.,~\cite{FPS-macbook19,CDDN21,CDN19,CDN18c,CDN20a,DFSY15,FPSV14,YS10}.

One of the basic tasks for mobile robots, intended as points in the plane, is certainly the requirement to achieve a placement so as no three of them are collinear. Furthermore, during the whole process, no two robots must occupy the same position concurrently, i.e., \emph{collisions} must be avoided.
This is known as the \textsc{Mutual Visibility} problem. The idea is that, if three robots are collinear, the one in the middle may obstruct the reciprocal visibility of the other two. 

Mutual Visibility has been largely investigated in recent years in many forms, subject to different assumptions. 
One main distinction within the Look-Compute-Move model concerns the level of synchronicity assumed among robots.
Robots are assumed to be synchronous~\cite{B20}, i.e., they are always all active and perform each computational cycle within the same amount of time; semi-synchronous~\cite{BM19,DFCPSV17,DFPSV14,SBM15}, i.e., robots are not always all active but all active robots always perform their computational cycle within a same amount of time, after which a new subset of robots can be activated; asynchronous~\cite{DFCPSV17,SBM15,ABKS18,BCM19,BM17,PAS21,SVT21}, i.e., each robot can be activated at any time and the duration of its computational cycle is finite but unknown.
Robots are generally endowed with visible lights of various colors useful to encode some information (to be maintained across different computational cycles and/or communicated to other robots), whereas in~\cite{DFPSV14} robots are considered completely oblivious, i.e., without any memory about past events.
Usually, robots are considered as points in the plane but in~\cite{PSA19}, where robots are considered ``fat'', i.e., occupying some space modeled as disks in the plane.
Furthermore, instead of moving freely in the Euclidean plane, in~\cite{ABKS18,SVT21} robots are constrained to move along the edges of a graph embedded in the plane and still the mutual visibility is defined according to the collinearity of the robots in the plane.

In this paper, we study the  \textsc{Geodesic Mutual Visibility} problem ($\GMV$, for short): 
starting from a configuration composed of robots located on distinct vertices of an arbitrary graph, within finite time the robots must reach, without collisions, a configuration where they all are in \emph{geodesic mutual visibility}. Robots are in geodesic mutual visibility if they are pairwise mutually visible, and two robots on a graph are mutually visible if there is a shortest path (i.e., a ``geodesic'') between them along which no other robots reside. This problem has been introduced in~\cite{CDDN23b} and can be thought of as a possible counterpart to the \textsc{Mutual Visibility} for robots moving in a discrete environment.

While this concept is interesting by itself, its study is motivated by the fact that robots, after reaching a  $\GMV$ condition, e.g., can communicate in an efficient and ``confidential'' way, by exchanging messages through the vertices of the graph that do not pass through vertices occupied by other robots or can reach any other robot along a shortest path without collisions. Concerning the last motivation, in~\cite{APS18}, it is studied the
{\sc Complete Visitability}  problem of repositioning a given number of robots on the vertices of a graph so that each robot has a path to all others without
visiting an intermediate vertex occupied by any other robot.
In that work, the required paths are not shortest paths and the studied graphs are restricted to infinite square and hexagonal grids, both embedded in the Euclidean plane.

The property of mutual visibility at the basis of $\GMV$ has been investigated in~\cite{D22} from a purely theoretical-graphic point of view: the goal is to understand how many robots, at most, can potentially be placed inside of a graph $G$ keeping the mutual visibility relation true. Such a maximum number of robots has been denoted by $\mu(G)$. In a general graph $G$, it turns out to be NP-complete to compute $\mu(G)$, whereas it has been shown that there are exact formulas for special graph classes like paths, cycles, trees, block graphs, co-graphs, and grids~\cite{D22,CiceroneDK23}. 
For instance, within a path $P$,  at most two robots can be placed, i.e., $\mu(P)=2$, whereas for a ring $R$, $\mu(R)=3$. In a finite square grid $\GSmn$ of $N>3$ rows and $M>3$ columns, $\mu(G)=2\min\{M,N\}$, whereas for a tree $T$, it has been proven that $\mu(T)=\ell(T)$, with $\ell(T)$ being the number of leaves of $T$.

\subsection{Results}
After recalling the problem of achieving $\GMV$ starting from a configuration of robots disposed on general graphs, we focus on square grids. The relevance of studying grids is certainly motivated by their peculiarity in representing a discretization of the Euclidean plane. Robots are assumed to have no explicit means of communication or memory of past events (we consider oblivious robots without lights). Hence, the movement of a robot does rely only on local computations on the basis of the snapshot acquired in the \Look phase.  Furthermore, in order to approach the problem, we make use of the methodology proposed in~\cite{CDN21a} that helps in formalizing the resolution algorithms as well as the related correctness proofs.

When studying $\GMV$ on square grids embedded in the plane, we add the further requirement to obtain a placement of the robots so as that the final minimum bounding rectangle enclosing all the robots is of minimum area. This area-constrained version of $\GMV$ is denoted as $\GMVopt$. We first solve $\GMVopt$ on \emph{finite} square grids embedded in the plane, and then provide relevant intuitions for extending the results to \emph{infinite} grids. In particular, we provide time-optimal algorithms that are able to solve $\GMVopt$ in both finite and infinite grid graphs. These algorithms work for synchronous robots endowed with chirality (i.e., a common handedness).


\subsection{Outline}
The rest of the paper is organized as follows.
Section~\ref{sec:model} introduces the robot model we have adopted. Section~\ref{sec:problem} formalizes the $\GMV$ problem and revises a resolution methodology to approach problems within the Look-Compute-Move context.
Section~\ref{sec:grids} deals with $\GMVopt$ on grids. It starts with some notation specific to the grid case, and then the resolution algorithm along with its correctness proof is intuitively and formally provided according to the recalled methodology. The section terminates with a description of the extension of the algorithm to deal with infinite grids. Section \ref{sec:concl} concludes the paper, posing possible future research directions.

\section{Robot model}
\label{sec:model}
%
Robots are modeled according to $\mathcal{OBLOT}$ (e.g., see~\cite{FPS19} for a survey), one of the classical theoretical models for swarm robotics. In this model, robots are computational entities that can move in some environment (a graph in our case) and can be characterized according to a large spectrum of settings. Each setting is defined by specific choices among a range of possibilities, with respect to a fundamental component - time synchronization - as well as other important elements, like memory, orientation and mobility. We assume such settings at minimum as follows:
\begin{itemize}
\item \emph{Anonymous}: no unique identifiers;
\item \emph{Autonomous}: no centralized control;
\item \emph{Dimensionless}: no occupancy constraints, no volume, modeled as entities located on vertices of a graph;
\item \emph{Oblivious}: no memory of past events;
\item \emph{Homogeneous}: they all execute the same \emph{deterministic}\footnote{No randomization features are allowed.} algorithm;
\item \emph{Silent}: no means of direct communication;
\item \emph{Disoriented}: no common coordinate system.
\end{itemize}
Each robot in the system has sensory capabilities allowing it to determine the location of other robots in the graph, relative to its own location.
Each robot refers in fact to a \emph{Local Coordinate System} (LCS) that might be different from robot to robot. Each robot follows an identical algorithm that is pre-programmed into the robot. 
The behaviour of each robot can be described according to the sequence of four states: \Wait, \Look, \Compute, and \Move. Such states form a computational cycle (or briefly a cycle) of a robot.
\begin{enumerate}
\item \Wait. The robot is idle. A robot cannot stay indefinitely idle;
\item  \Look. The robot observes the environment by activating its sensors which will return a snapshot of the positions of all other robots with respect to its own LCS. Each robot is viewed as a point; 
\item  \Compute. The robot performs a local computation according to a deterministic algorithm $\A$ (we also say that the robot executes $\A$). The algorithm is the same for all robots, and the result of the \Compute phase is a destination point. 
Actually, for robots on graphs, the result of this phase either is the vertex where the robot currently resides 
or it is a vertex among those at one hop distance 
(i.e., at most one edge can be traversed);
\item  \Move. If the destination point is the current vertex where $r$ resides, $r$ performs a $\nil$ movement (i.e., it does not move); otherwise, it moves to the adjacent vertex selected.
\end{enumerate}
When a robot is in \Wait, we say it is \emph{inactive}, otherwise it is \emph{active}. In the literature, the computational cycle is simply referred to as the \Look-\Compute-\Move (LCM) cycle, as during the \Wait phase a robot is inactive. 

Since robots are oblivious, they have no memory of past events. This implies that the \Compute phase is based only on what is determined in their current cycle (in particular, from the snapshot acquired in the current \Look phase). A data structure containing all the information elaborated from the current snapshot represents what later is called the \emph{view} of a robot. Since each robot refers to its own LCS, the view cannot exploit absolute orienteering but it is based on relative positions of robots. 

Concerning the movements, in the graph environment moves are always considered as instantaneous. This results in always perceiving robots on vertices and never on edges during \Look phases. Hence, robots cannot be seen while moving, but only at the moment they may start moving or when they arrived. 
Two or more robots can move toward the same vertex at the same time, thus creating what it called a \emph{multiplicity} (i.e., a vertex occupied by more than one robot). When undesired, a multiplicity is usually referred to as a \emph{collision}.

In the literature, different characterizations of the environment have been considered according to whether robots are fully-synchronous, semi-synchronous, or asynchronous (cf.~\cite{CDN21,FPS19}). These synchronization models are 
defined as follows:

\begin{itemize}
\item \emph{Fully-Synchronous} (\fsync): All robots are always active, continuously executing in a synchronized way their \LCM-cycles. Hence the time can be logically divided into
global rounds. In each round, all the robots obtain a snapshot of the environment, compute on the basis of the obtained snapshot and perform their computed move;
\item \emph{Semi-Synchronous} (\ssync): robots are synchronized as in \fsync but
not all robots are necessarily activated during a \LCM-cycle; 
\item \emph{Asynchronous} (\async): Robots are activated independently,
and the duration of each phase is finite but
unpredictable. As a result, robots do not have a common notion of time.
\end{itemize}

In \async, the amount of time to complete a full \LCM-cycle is assumed to be finite but unpredictable. Moreover, in the \ssync and \async cases, it is usually assumed the existence of an \emph{adversary} which determines the computational cycle's timing. Such timing is assumed to be \emph{fair}, that is, each robot performs its \LCM-cycle within finite time and infinitely often. Without such an assumption the adversary may prevent some robots from ever moving.

It is worth remarking that the three synchronization schedulers induce the following hierarchy (see~\cite{DDFN18}): \fsync robots are more powerful (i.e., they can solve more tasks) than \ssync robots, that in turn are more powerful than \async robots. This simply follows by observing that the adversary can control more parameters in \async than in \ssync, and more in \ssync than in \fsync. In other words, protocols designed for \async robots also work for \ssync and \fsync robots. On the contrary, any impossibility result proved for \fsync robots also holds for \ssync, and \async robots.  

Whatever the assumed scheduler is, the activations of the robots according to any algorithm $\A$ determine a sequence of specific time instants $t_0 < t_1 < t_2 < \ldots$ during which at least one robot is activated. Apart from the \async case where the notion of time is not shared by robots, for the other types of schedulers robots are synchronized. In the \fsync case, each robot is active at each time unit. In the \ssync, we assume that at least one robot is active at each time $t$.
If $C(t)$ denotes the configuration observed by some robots at time $t$ during their \Look phase, then an \emph{execution} of  $\A$ from an initial configuration $C$ is a sequence of configurations $\Ex: C(t_0),C(t_1),\ldots$, where $C(t_0)=C$ and $C(t_{i+1})$ is obtained from $C(t_i)$ by moving at least one robot (which is active at time $t_i$) according to the result of the \Compute phase as implemented by $\A$. 
Note that, in \ssync or \async there exists more than one execution of $\A$ from $C(t_0)$ depending on the activation of the robots or the duration of the phases, whereas in \fsync the execution is unique as it always involves all robots in all time instants.

\section{Problem formulation and resolution methodology}\label{sec:problem}

The topology where robots are placed is represented by a simple and connected graph $G=(V,E)$. A function $\lambda: V\to \mathbb{N}$ gives the number of robots on each vertex of $G$, and we call $C=(G,\lambda)$ a \emph{configuration} whenever $\sum_{v\in V} \lambda(v)$ is bounded and greater than zero. In this paper, we introduce the \textsc{Geodesic Mutual Visibility} ($\GMV$, for short) problem: \\

\problemnr{$\GMV$}{prob:gmv}{%
A configuration $C=(G,\lambda)$ in which each robot lies on a different vertex of a graph $G$.}%
{Design a deterministic distributed algorithm working under the \LCM model that, starting from $C$, brings all robots on distinct vertices -- without generating collisions -- in order to obtain the geodesic mutual visibility, that is there is a geodesic between any pair of robots where no other robots reside.%
} 

Since the definition of mutual visibility requires that robots are located in distinct vertices, then the above definition requires that any possible solving algorithm does not create collisions. In fact, as the robots are anonymous and homogeneous, regardless of the synchronicity model, the adversary will be able to keep the multiplicity unchanged and no algorithm will ever be able to separate the robots. It is worth remarking that this is a special case with respect to the general situation in which a configuration contains \emph{equivalent} robots. The next paragraph provides a formal definition of such an equivalence relationship.


\subsection{Symmetric configurations}\label{ssec:symmetries}
Two undirected graphs $G=(V,E)$ and $G'=(V',E')$ are \emph{isomorphic} if there is a bijection $\varphi$ from $V$ to $V'$ such that $\{u,v\} \in E$ iff $\{\varphi(u),\varphi(v)\} \in E'$. An \emph{automorphism} on a graph $G$ is an isomorphism from $G$ to itself, that is a permutation of the vertices of $G$ that maps edges to edges and non-edges to non-edges. The set of all automorphisms of $G$ forms a group called \emph{automorphism group} of $G$ and is denoted by $\Aut(G)$. If $|\Aut(G)|=1$, that is $G$ admits only the identity automorphism, then the graph $G$ is 
called
\emph{asymmetric}, otherwise it is 
called 
\emph{symmetric}. Two vertices $u$, $v\in V$ are \emph{equivalent} if there exists an automorphism $\varphi\in \Aut(G)$ such that $\varphi(u)=v$.

The concept of isomorphism can be extended to configurations in a natural way: two configurations $C=(G,\lambda)$ and $C'=(G',\lambda')$ are isomorphic if $G$ and $G'$ are isomorphic via a bijection $\varphi$ and $\lambda(v)=\lambda'(\varphi(v))$ for each vertex $v$ in $G$.  An \emph{automorphism} on $C$ is an isomorphism from $C$ to itself and the set of all automorphisms of $C$ forms a group that we call \emph{automorphism group} of $C$ and denote by $\Aut(C)$. Analogously to graphs, if $|\Aut(C)|=1$, we say that the configuration $C$ is \emph{asymmetric}, otherwise it is \emph{symmetric}. In a configuration $C$, two robots $r_1$ and $r_2$, respectively located on distinct vertices $v_{r_1}$ and $v_{r_2}$,  are \emph{equivalent} if there exists $\varphi\in \Aut(C)$ such that $v_{r_2}=\varphi(v_{r_1})$. Note that $\lambda(v_{r_1})=\lambda(v_{r_2})$ whenever $v_{r_1}$ and $v_{r_2}$ are equivalent.

From an algorithmic point of view, it is important to remark that when $\varphi\in \Aut(C)$ makes the elements of $V'\subseteq V$ pairwise equivalent, then a robot $r_1$ cannot distinguish its position $v_{r_1}\in V'$ from that of a robot $r_2$ located at vertex $v_{r_2} = \varphi(v_{r_1})\in V'$.  As a consequence, no algorithm can distinguish between two equivalent robots, and then it cannot avoid the adversary activates two equivalent robots at the same time and that they perform the same move simultaneously.

\subsection{Methodology}\label{sec:methodology}
The algorithms proposed in this paper are designed according to the methodology proposed in~\cite{CDN21a}. Assume that an algorithm $\A$ must be designed to resolve a generic problem $\Prob$. Here we briefly summarize how $\A$ can be designed according to that methodology.

In general, a single robot has rather weak capabilities with respect to $\Prob$ it is asked to solve along with other robots (we recall that robots have no direct means of communication). For this reason, $\A$ should be based on a preliminary decomposition approach:  $\Prob$ should be divided into a set of sub-problems so that each sub-problem is simple enough to be thought of as a ``task'' to be performed by (a subset of) robots. This subdivision could require several steps before obtaining the definition of such simple tasks, thus generating a sort of hierarchical structure. Assume now that $\Prob$ is decomposed into simple tasks $T_1,T_2,\ldots,T_k$, where one of them is the \emph{terminal} one, i.e. the robots recognize that the current configuration is the one in which $\Prob$ is solved and do not make any moves.

According to the \LCM model, during the \Compute phase, each robot must be able to recognize the task to be performed just according to the configuration perceived during the \Look phase. 
This recognition can be performed by providing $\A$ with a \emph{predicate} $P_i$ for each task $T_i$. Given the perceived configuration, the predicate $P_i$ that results to be true reveals to robots that the corresponding task $T_i$ is the task to be performed. With predicates $P_i$ well-formed, algorithm $\A$ could be used in the \Compute phase as follows: \textit{-- if a robot $r$ executing algorithm $\A$ detects that   predicate $P_i$ holds, then $r$ simply performs a move $m_i$ associated with task $T_i$}. In order to make this approach valid, the well-formed predicates must guarantee the following  properties:
\begin{description}
\item[$\Prop_1$:]
each $P_i$ must be computable on the configuration $C$ perceived in each \Look phase; 
\item[$\Prop_2$:]
$P_i \wedge P_j =\false$, for each $i\neq j$; this property allows robots to exactly recognize the task to be performed;
\item[$\Prop_3$:]
for each possible perceived configuration $C$, there must exist a predicate $P_i$ evaluated true. 
\end{description}

Concerning the definition of the predicates, it is reasonable to assume that each task $T_i$ requires some \emph{precondition} to be verified. Hence, in general, to define the predicates we need:
\begin{itemize}
\item
\emph{basic variables} that capture metric/topological/numerical/ordinal aspects of the input configuration which are relevant for the used strategy and that can be evaluated by each robot on the basis of its view;
\item
\emph{composed variables} that express the preconditions of each task $T_i$.
\end{itemize}
If we assume that $\pre_i$ is the composed variable that represents the preconditions of $P_i$, for each $1\le i\le k$, then predicate $P_i$ can be defined as follow:
\begin{equation}\label{eq:predicates}
	P_i = \pre_i \wedge \neg ( \pre_{i+1} \vee \pre_{i+2} \vee \ldots \vee \pre_k )
\end{equation} 
This definition ensures that any predicate fulfils Property $\Prop_2$ (it is  directly implied by Equation~\ref{eq:predicates}). 

Consider now an execution of $\A$, and assume that a task $T_i$ is performed with respect to the current configuration $C$. If $\A$ transforms $C$ into $C'$ and this new configuration has to be assigned the task $T_j$, then we say that $\A$ can generate a transition from $T_i$ to $T_j$. The set of all possible transitions of $\A$ determines a directed graph called \emph{transition graph}. Of course, the terminal task among $T_1,T_2,\ldots,T_k$ must be a sink node in the transition graph.

According to the proposed methodology, in~\cite{CDN21a} it is shown that the correctness of $\A$ can be obtained by proving that all the following properties hold:

\begin{itemize}
\item[$\h_1$:]
for each task $T_i$, the tasks reachable from $T_i$ by means of transitions are exactly those represented in the transition graph (i.e., the transition graph is correct);
\item[$\h_2$:]
possible cycles in the transition graph (including self-loops) must be performed a finite number of times -- apart for the self-loop induced by a  terminal task;
\item[$\h_3$:]
unsolvable configurations are not generated by $\A$ (with respect to $\GMV$, for instance, this means that $\A$ does not generate multiplicities, i.e., it is collision-free).
\end{itemize}

\section{Solving GMV on square grids}\label{sec:grids}

In this section, we solve $\GMV$ for robots moving on finite or infinite square grids embedded in the plane. Moreover, we add the further requirement to obtain a placement of the robots so as that the final minimum bounding rectangle enclosing all the robots is of minimum area. This area-constrained $\GMV$ problem is denoted as $\GMVopt$. Such a requirement avoids `trivial' solutions -- in terms of feasibility, especially in the case of infinite grids. In fact, by aligning all the robots along a diagonal, $\GMV$ would be solved. In a finite grid $\GSmn$, instead, this is not always possible, depending on the number of robots. Hence, different approaches are required anyway. Moreover, when the number of robots is exactly $\mu(G)$, then the minimum area constraint is forced.

While considering synchronous robots endowed with chirality (i.e., they share a common handedness), we provide time-optimal algorithms solving $\GMVopt$ in both finite and infinite grid graphs. 

\subsection{Preliminary concepts and notation}
\begin{figure}[t]
\centering
\includegraphics[scale=0.60]{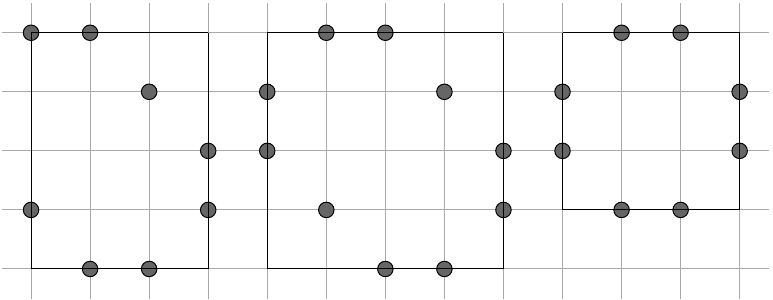} 
\caption{\small Examples of $\mbr(R)$.}
\label{fig:mbr}
\end{figure}

Given a graph $G$, let $d(u, v)$ be the distance in $G$ between two vertices $u$, $v\in V$ in terms of the minimum number of edges traversed. We extend the notion of distance to robots: given $r_i,r_j\in R$, $d(r_i, r_j)$ represents the distance between the vertices in which the robots reside. $D(r)$ denotes the sum of distances of $r \in C$ from any other robot, that is $D(r) =\sum_{r_i\in C}{d(r,r_i)}$. 
A square tessellation of the Euclidean plane is the covering of the plane using squares of side length 1, called tiles, with no overlaps and in which the corners of squares are identically arranged. Let $S$ be the infinite lattice formed by the vertices of the square tessellation. The graph called infinite \textbf{grid graph}, and denoted by $\GS$, is such that its vertices are the points in $S$ and its edges connect vertices that are distance 1 apart. In this section, $\GSmn$ denotes a finite grid graph formed by $M\cdot N$ vertices (i.e., informally generated by  $M$ ``rows'' and $N$ ``columns''). 
By $\mbr(R)$, we denote the \textbf{minimum bounding rectangle} of $R$, that is the smallest rectangle (with sides parallel to the edges of $\GSmn$) enclosing all the robots (cf. Figure~\ref{fig:mbr}). Note that $\mbr(R)$ is unique. By $c(R)$, we denote the center of $\mbr(R)$. 

\block{Symmetric configurations}\label{ssec:symmetries-grids}
As chirality is assumed, then 
the only possible symmetries that can occur in our setting are rotations of 90 or 180 degrees. A rotation is defined by a center $c$ and a minimum angle of rotation $\alpha\in \{90,180,360\}$ working as follows: if the configuration is rotated around $c$ by an angle $\alpha$, then a configuration coincident with itself is obtained. The \textbf{order} of a configuration is given by $360/\alpha$. A configuration is \textbf{rotational} if its order is 2 or 4. The \textbf{symmetricity} of a configuration $C$, denoted as $\rho(C)$, is equal to its order, unless its center is occupied by one robot, in which case $\rho(C)=1$. Clearly, any asymmetric configuration $C$ implies $\rho(C)=1$.

\begin{figure}[h]
\centering
\includegraphics[scale=0.90]{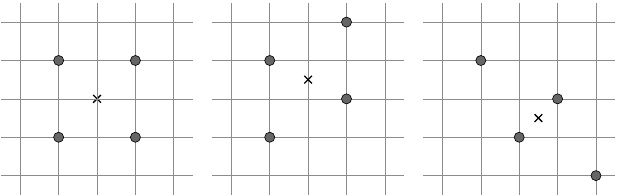}     
\caption{\small Examples for the notion of center of three rotational configurations: in order, $\tc(C_1)=1$, $\tc(C_2)=2$, and $\tc(C_3)=3$.}
\label{fig:centers}
\end{figure}

The \textbf{type of center} of a rotational configuration $C$ is denoted by $\tc(C)$ and is equal to 1, 2, or 3 according to whether the center of rotation is on a vertex, on a median point of an edge, or on the center of a square of the tessellation forming a grid, respectively (cf. Figure~\ref{fig:centers}).

\begin{figure}[t]
\centering
\def\svgwidth{\columnwidth}
\large\scalebox{.75}{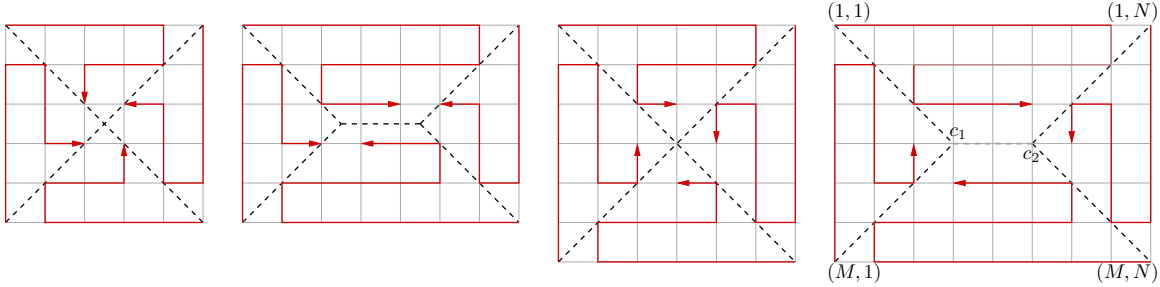}     
\caption{\small Examples of \specialpaths with respect to different configurations.}
\label{fig:snakes}
\end{figure}

%
\block{The view of robots} 
In $\algo$, robots encode the perceived configuration into a binary string called  \textbf{lexicographically smallest string} and denoted as $\LSS(R)$ (cf.~\cite{DDKN12,CDDN23}). To define how robots compute the string, we first analyze the case in which $\mbr(R)$ is a square: the grid enclosed by $\mbr(R)$ is analyzed row by row or column by column starting from a corner and proceeding clockwise, and 1 or 0 corresponds to the presence or the absence, respectively, of a robot for each encountered vertex. This produces a string assigned to the starting corner, and four strings in total are generated. If $\mbr(R)$ is a rectangle, then the approach is restricted to the two strings generated along the smallest sides. The lexicographically smallest string is the $\LSS(R)$. Note that, if two strings obtained from opposite corners along opposite directions are equal, then the configuration is rotational, otherwise it is asymmetric. 
The robot(s) with \textbf{minimum view} is the one with minimum position in $\LSS(R)$. The first three configurations shown in Figure~\ref{fig:centers} can be also used for providing examples about the view. 
In particular: 
in the first case,  we have $\rho(C)=1$ and $\LSS(R)$ = 0110 1001 1000 0100 0011; 
in the second case, we have $\rho(C)=2$ and $\LSS(R)$ = 00110 01001 10001 10010 01100; 
in the last case,   we have $\rho(C)=4$ and $\LSS(R)$ = 0110 1001 1001 0110. 
\block{Regions}
Our algorithms assume that robots are assigned to \textbf{regions} of $\mbr(R)$ as follows (cf. Figure~\ref{fig:snakes}). If $\mbr(R)$ is a square, the four regions are those obtained by drawing the two diagonals of $\mbr(R)$ that meet at $c(R)$. If $\mbr(R)$ is a rectangle, then from each of the vertices positioned on the shorter side of $\mbr(R)$ starts a line at 45 degrees toward the interior of $mbr(R)$ - these two pairs of lines meet at two points (say $c_1(R)$ and $c_2(R)$) which are then joined by a segment.

In each of the four regions, it is possible to define a \textbf{\specialpath} that starts from a corner $v$ and goes along most of the vertices in the region. To simplify the description of such a path, assume that $\mbr(R)$ coincides with a sub-grid with $M$ rows and $N$ columns, and the vertices are denoted as $(i,j)$, with $1\le i\le M$ and $1\le j\le N$. The \specialpath that starts at $(1,1)$ is made of a sequence of ``traits'' defined as follows: the first trait is $(1,1), (1,2), \ldots, (1,N-1)$, the second is $(2,N-1), (2,N-2), \ldots, (2,3)$, the third is $(3,3), (3,4), \ldots, (3,N-3)$, and so on. This process ends after $\lfloor \min\{M,N\}/2 \rfloor$ traits are formed in each region, and the \specialpath is obtained by composing, in order, the traits defined in each region (see the red lines in Figure~\ref{fig:snakes}).

\subsection{An algorithm for $\GMVopt$}\label{sec:algo-grid}
%
In this section, we present a resolution algorithm for the $\GMVopt$ problem, when considering $n\ge 7$ fully synchronous robots endowed with chirality and moving on a finite grid graph $\GSmn$ with $M,N\ge \lceil \frac n 2 \rceil$ rows and columns. Note that the constraints on the number of rows and columns depend on the fact that on each row (or column) it is possible to place at most two robots, otherwise the outermost robots on the row (or column) are not in mutual visibility. Concerning the number of robots, we omit the cases with $n<7$ as they require just tedious and specific arguments that cannot be generalized. Hence, we prefer to cut them out of the discussion, even though they can be solved. 

Our approach is to first design a specific algorithm $\algoAsym$ that solves $\GMVopt$ only for \textbf{asymmetric configurations}. Later, we will describe (1)  how $\algoAsym$ can be extended to a general algorithm $\algo$ that also handles symmetric configurations, and (2)  how, in turn, $\algo$ can be modified into an algorithm $\algoinfty$ that solves the same problem for each input configuration defined on infinite grids.

\begin{figure}[t]
\centering
\includegraphics[scale=0.7]{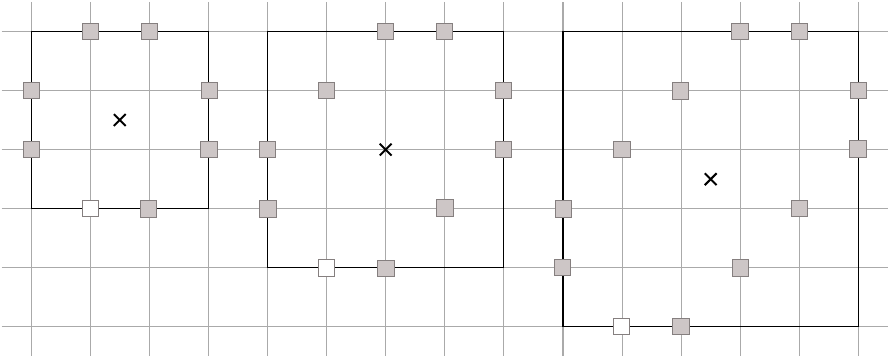} 
\caption{\small Patterns $F$ for asymmetric input configurations with $n=8,10,12$ robots. For $n=7,9,11$, the position represented in white is not considered in $F$.
}
\label{fig:pattern}
\end{figure}

\block{The pattern formation approach} 
$\algoAsym$ follows the ``pattern formation'' approach. In the general pattern formation problem, robots belonging to an initial configuration $C$ are required to arrange themselves in order to form a configuration $F$ which is provided as input. In~\cite{CDN21b,SY99}, it is shown that $F$ can be formed if and only if $\rho(C)$ divides $\rho(F)$. Hence, here we show some patterns that can be provided as input to $\algoAsym$ so that:
\begin{enumerate}
\item $\rho(C)$ divides $\rho(F)$;
\item if $\rho(C)\in \{2,4\}$ then $\tc(C) = \tc(F)$;
\item the positions specified by $F$ solve $\GMVopt$.
\end{enumerate}

The first requirement trivially holds since we are assuming that $C$ is asymmetric and hence $\rho(C)=1$. The second is required since the center of symmetric configurations is an invariant for synchronous robots. Concerning the last requirement, in Figure~\ref{fig:pattern} we show some examples for $F$ when $7\le n\le 12$. In~\cite{D22}, it is shown how $F$ is defined for any $n$ and it is also proved that the elements in these patterns always solve $\GMV$ for the grid $\GSmn$. Finally, since in $F$ there are two robots per row and per column, and since in $\mbr(F)$ all the rows and columns are occupied (for $n$ even), it can be easily observed that $F$ solves $\GMVopt$. 

\block{High level description of the algorithm}\label{alg-highlevel}
The algorithm is designed according to the methodology recalled in Section~\ref{sec:methodology} that allows dividing the problem $\GMVopt$ into a set of sub-problems that are simple enough to be thought as ``tasks'' to be performed by (a subset of) robots.

As a first sub-problem, the algorithm $\algoAsym$ selects a single robot, called guard $r_g$, to occupy a corner of the grid $\GSmn$. As robots are disoriented (only sharing chirality), the positioning of the guard allows the creation of a common reference system used by robots in the successive stages of the algorithm. Given chirality, the position of $r_g$ allows robots to identify and enumerate rows and columns. $r_g$ is not moved until the final stage of the algorithm and guarantees that the configuration $C$ is kept asymmetric during the movements of the other robots. Given the common reference system, all robots agree on the embedding of the pattern $F$, which is realized by placing the corner of $F$ with the maximum view in correspondence with the corner of $\GSmn$ in which $r_g$ resides. This sub-problem  is solved by tasks $T_{1a}, T_{1b}$, or $T_{1c}$.
In task $T_2$, the algorithm moves the robots so as to obtain the suitable number of robots for each row according to pattern $F$, that is, two robots per row. The only exception comes when $n$ is odd, in which case the last row will require just one robot. During task $T_3$, robots move toward their final target along rows, except for $r_g$. When $T_3$ ends, $n-1$ robots are in place according to the final pattern $F$. During task $T_4$, $r_g$ moves from the corner of $\GSmn$ toward its final target, placed on a neighbouring vertex, hence leading to the final configuration in one step.

\subsection {Detailed description of the tasks}\label{sec:tasks-grid}
In this section, we provide all the necessary details for each of the designed tasks. 

\block{Task $T_1$} Here the goal is to select a single robot $r_g$ to occupy a corner 
of the grid $\GSmn$. This task is divided into three sub-tasks based on the number of robots occupying the perimeter -- and in particular the corners, of $\GSmn$. Let $\RS$ be the number of robots on the sides of $G$, and let $\RV$ be the number of robots on the corners of $G$.

Task $T_{1a}$ starts when there are no robots on the perimeter of $G$ and selects the robot $r_g$ such that $D(r)$ is maximum, with $r$ of minimum view in case of ties. The planned move is $m_{1a}$: \textit{$r_g$ moves toward the closest side of $G$.} At the end of the task, $r_g$ is on the perimeter of $\GSmn$.

Task $T_{1b}$ activates when the following precondition holds: 
$$\small \pre_{1b}  \equiv \RS \ge 1 \wedge \RV=0.$$ 
In this case, there is more than one robot on the perimeter of $\GSmn$ but none on corners. The task selects the robot $r_g$ located on a side of $G$ closest to a corner of $\GSmn$, with the minimum view in case of ties, to move toward a corner of $\GSmn$.
Move $m_{1_b}$ is defined as follows: \textit{$r_g$ moves toward the closest corner of $\GSmn$} -- arbitrarily chosen if more than one. At the end of task $T_{1b}$, a single robot $r_g$ occupies a corner of the grid $\GSmn$.

Task $T_{1c}$  activates when the following precondition holds:
$$\small \pre_{1c}  \equiv \RV > 1.$$
In this case, all the robots on the corners but one move away from the corners. The moves are specified by Algorithm~\ref{alg:serpentone}. This algorithm uses some additional definitions. In particular, a \specialpath is said \textbf{occupied} if there is a robot on its corner. A \specialpath is said to be \textbf{fully-occupied} if robots are placed on all its vertices. Given an occupied \specialpath $P$, a \specialsubpath is a fully occupied sub-path of $P$ starting from the corner of $P$. Finally, $\FO$ denotes the number of fully-occupied special-paths.

\begin{algorithm}[t]
\caption{MoveAlong~\specialpath}\label{alg:serpentone}
\begin{algorithmic}[1]
\begin{small}
 \REQUIRE a configuration $C$
     	
    \IF {$\FO=0$}\label{l:nonFully}
        
        \STATE Let $S$ be the occupied \specialpath whose first robot has the minimum view.
        \STATE \textbf{move}: all the robots on a \specialsubpath and not on $S$ move toward the neighbor vertex along the \specialpath.
      \ENDIF 
      
      \IF {$\FO=1$ }\label{l:Fully}
      \STATE Let $I$ be the fully-occupied  \specialpath
      \STATE \textbf{move}: all the robots on a \specialsubpath and not on $I$ move toward the neighbor vertex along the \specialpath
       \ENDIF
      \IF {$\FO=2$}\label{l:moreFully}
      \STATE \textbf{move}: the robot on a corner of $\GSmn$, with an empty neighbor, moves toward it.
     \ENDIF

\end{small}
\end{algorithmic} 
\end{algorithm}

At line~\ref{l:nonFully}, the algorithm checks if there are no fully-occupied \specialpaths. In this case, there are at least two occupied \specialpaths. The robot, occupying the corner, with minimum view, is elected as 
guard $r_g$. The move is designed to empty all the other corners of $\GSmn$ except for the one occupied by $r_g$. In each occupied \specialpaths, but the one to which $r_g$ belongs to, the robots on the corners, and those in front of them along the \specialpaths until the first empty vertex, move forward along the \specialpath.
At line~\ref{l:Fully}, there is exactly one fully-occupied special path. Therefore, robots on the fully-occupied \specialpath are kept still. Concerning the other occupied \specialpaths, the robots on corners, and those in front of them until the first empty vertex, move forward along the \specialpath.
At line~\ref{l:moreFully} there is more than one fully-occupied \specialpath. Actually, this condition can occur only for a $4\times 4$ grid $\GSmn$ with two fully-occupied \specialpaths located on two successive corners of $\GSmn$. Therefore, there is a single robot $r$, on a corner of $\GSmn$, with an empty neighbour. Then, $r$ moves toward that neighbour.

Note that, Algorithm~\ref{alg:serpentone} is designed so that, in a robot cycle, a configuration is obtained where exactly one corner of $\GSmn$ is occupied.

\block{Task $T_2$} In task $T_2$, the algorithm moves the robots to place the suitable number of robots for each row according to the pattern $F$, starting from the first row, while possible spare rows remain empty. At the end of the task, for each row corresponding to those of the pattern $F$, there are two robots, except when the number of robots $n$ is odd, in which case in the last row is placed a single robot. The position of $r_g$ allows robots to identify the embedding of $F$ and hence the corresponding rows and columns. We assume, without loss of generality, that $r_g$ is positioned on the upper-right corner of $\GSmn$. $r_g$ identifies the first row. In this task, we define $c(r)$ and $l(r)$ as the column and the row, respectively, where robot $r$ resides. Columns are numbered from left to right, therefore $l(r_g)=1$ and $c(r_g)=N$. Let $t_l$ be the number of targets on row $l$ in $F$, let  $(t_1, t_2,\dots,t_M)$ be the vector of the number of targets, and let $(n_1, n_2, \ldots,n_M)$ be the number of robots on each of the $M$ rows of $\GSmn$. 

For each row $l$, the algorithm computes the number of exceeding robots above and below $l$ with respect to the number of targets, to determine the number of robots that need to leave row $l$. Given a row $l$, let $R_l$ be the number of robots on rows from 1 to $l-1$, and let $R’_l$ be the number of robots on rows from $l+1$ to $M$. Accordingly, let $T_l$ and $T’_l$ be the number of targets above and below the line $l$, respectively. We define the subtraction operation $\dotdiv$ between two natural numbers $a$ and $b$ as $a \dotdiv b=0$ if $a<b$, $a \dotdiv b=a - b$, otherwise. Concerning to the number of targets, given a row $l$, let $\Bl$ be the number of exceeding robots above $l$, $l$ included, and let $\Al$ be the number of exceeding robots below $l$, $l$ included. Formally, $\Bl = (R_l + n_l) \dotdiv (T_l + t_l)$ and $\Al = (R’_l + n_l) \dotdiv (T’_l + t_l)$.

Let $RD_l=n_l - (n_l \dotdiv \Bl)$ be the number of robots that must move downward and  $RU_l=n_l - (n_l \dotdiv \Al)$ be the number of robots that must move upward from row $l$. Task $T_2$ activates when precondition $\pre_2$ becomes true:
$$\small \pre_2  \equiv \RV=1 \wedge \exists \enspace l \in {1,\dots ,M} : \Bl\neq 0 \lor \Al\neq 0.$$
The precondition identifies the configuration in which the guard $r_g$ is placed on a corner of $\GSmn$ and there is at least a row in which there is an excess of robots.
We define \textbf{outermost} any robot that resides on the first or the last column of $\GSmn$. Let $U_l$ ($D_l$, resp.) be a set of robots on row $l$ chosen to move upward (downward, resp.) and let $U$ ($D$, resp.) be the list of sets $U_l$ ($D_l$, resp.) with $l \in \{1,\ldots, M\}$. The robots that move upward or downward are chosen as described in Algorithm~\ref{alg:selection}.

\begin{algorithm}[t]
\caption{SelectRobots}\label{alg:selection}
\begin{algorithmic}[1]
\begin{footnotesize}

 \REQUIRE $C'=(C \setminus r_g$) 

 	\STATE Let $U=\{U_1,U_2,\ldots,U_M\}$ be a list of empty sets
 	\STATE Let $D=\{D_1,D_2,\ldots,D_M\}$ be a list of empty sets

 	\FORALL {$l \in (1\ldots m)$} 
    \STATE	$\Bl \gets (R_l + n_l) \dotdiv (T_l + t_l)$ \label{l:Bl}
    \STATE $\Al \gets (R’_l + n_l) \dotdiv (T’_l + t_l)$
    \STATE $RD_l \gets n_l - (n_l \dotdiv \Bl)$
    \STATE $RU_l \gets n_l - (n_l \dotdiv \Al)$\label{l:RU} 
 	
    \IF {$M > \lceil n/2 \rceil$}
     \STATE Let $U_l$ be the set of $RU_l$ robots of row $l$ selected from right
     
     \STATE Let $D_l$ be the set of $RD_l$ robots of row $l$ selected from left

    \ELSE \label{l:rows=n/2}  
    \STATE Let $U_l$ be the set of $RU_l$ robots of row $l$ from right and not outermost
    \STATE Let $D_l$ be the set of $RD_l$ robots of row $l$ from left and not outermost

   \ENDIF
    \ENDFOR
    \STATE \textbf{if} $U_2=\{r\}$ and $l(r)=2$ and $c(r)=1$ \textbf{then} $U_2=\emptyset$\label{l:row2}

 \RETURN $U$, $D$ \label{l:return}
\end{footnotesize}
\end{algorithmic} 
\end{algorithm}

For each row $l$, at lines~\ref{l:Bl}--\ref{l:RU}, the algorithm computes the number of exceeding robots $\Bl$, $\Al$, and the number of robots that must leave the row $RD_l$ and $RU_l$. Then, it checks whether the number $M$ of rows of $\GSmn$ is greater than the number $k$ of rows of $F$. 
The algorithm selects $RD_l$ robots to move downward, starting from the first column, and $\Al$ robots to move upward, starting from the $N$-th column. 

Line~\ref{l:rows=n/2} corresponds to the case in which $M=k$, the algorithm selects $RD_l$ robots to move downward, starting from the second column and $RU_l$ robots to move upward, starting from the $N-1$ column. This avoids the selection of robots that may move in one of the corners of $\GSmn$.
At line~\ref{l:row2}, the algorithm checks if a robot $r$ selected to move upward on row 2, occupies vertex $(2,1)$. In the positive case, $r$ is removed from $U_2$. This avoids $r$ to move to a corner of $\GSmn$. 
At line~\ref{l:return}, the algorithm returns the sets $U$ of robots chosen to move upward for each row,  and the sets $D$ of robots chosen to move downward.
Given a robot $r$ on a row $l$, let $AlignedUp$ be the Boolean variable that is true when there exists another robot $r'$ such that $(U_{l+1}=\{r'\} \textrm{ and } c(r)=c(r'))$ holds, and $AlignedDown$ be the Boolean variable that is true when there exists another robot $r''$ such that $(D_{l-1}=\{r''\}  \textrm{ and } c(r)=c(r''))$ holds.
Let $t(r)$ be the target of a robot $r$ defined as follows:
\begin{equation}
\small
\label{formula}
t(r) =
\begin{cases}
(l(r)+1, c(r))  & \textrm{ if } r \in D_l  \\
(l(r)-1, c(r))  & \textrm{ if } r \in U_l  \\
(l(r),c(r)-1)   & \textrm{ if } (AlignedUp \textrm{ or }  AlignedDown) \textrm{ and } c(r)\ge N/2 \\
		     
(l(r),c(r)+1)   & \textrm{ if } (AlignedUp \textrm{ or }  AlignedDown) \textrm{ and } c(r) < N/2 \\
(2,2)     & \textrm{ if } RU_2=1 \textrm{ and } \exists! \textrm{ r on } l_2 \textrm{ | } c(r)=1 \textrm{ and } l(r)=2  \\
(l(r), c(r) )     &  ~otherwise    
\end{cases}
\end{equation}

The first two cases reported in the definition of Equation~(\ref{formula}) 
identify the target of robot $r$ when is selected to move downward (upward, resp.). The target of $r$ is one row below (above, resp.) its current position and on the same column. The third and the fourth cases refer to the occurrence in which there is a robot $r_1$, positioned in the same column of $r$, that is selected to move upward or downward. Then, the target of $r$ is on a neighbouring vertex, on the same row, closer to the center of $\GSmn$.
The fifth case reports the target of a robot $r$ when positioned on the second row and first column, and one robot is required to move on the first row. To avoid occupying a corner of $\GSmn$, the target of $r$ is the neighbouring vertex to $r$ on its same row. In all other cases, the target of a robot $r$ is its current position. Robots move according to Algorithm~\ref{alg:move}.

\begin{algorithm}[t]
\caption{MoveRobot}\label{alg:move}
\begin{algorithmic}[1]
\begin{small}
 \REQUIRE a configuration $C$, guard $r_g$
 
 \STATE $U$, $D$ = SelectRobots($C \setminus r_g$)\label{l:selectRobots}
 \FORALL {robots $r$}\label{l:computeTargets1}
 \STATE Compute $t(r)$\label{l:computeTargets2}
 \ENDFOR
 
    \IF {$r \notin  U_{l(r)}$ or  $\forall$  $r_1, r_2$, $t(r_1)\neq t(r_2)$}\label{l:checkConflicts}

        \STATE move to $t(r)$ \label{l:move}
      \ENDIF 
\end{small}
\end{algorithmic} 
\end{algorithm}

\begin{figure}[t]
   \centering
   \def\svgwidth{\columnwidth}
   \large\scalebox{.65}{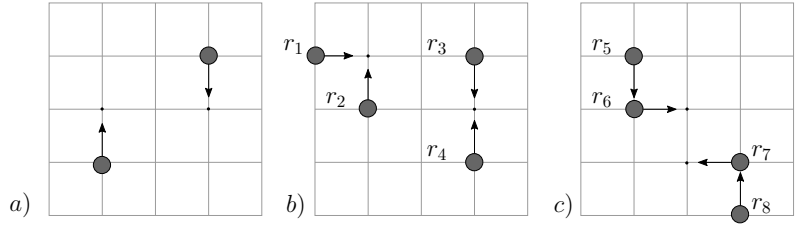}
    \caption{\small The three possible movement combinations as described in task $T_2$. Grey circles represent robots, arrows represent the direction of movements, and small dots are robot targets.
}
\label{fig:movements}
\end{figure}

Each robot runs Algorithm~\ref{alg:move} independently.
At line~\ref{l:selectRobots}, a robot calls procedure SelectRobots on $C'=\{C \setminus r_g\}$ and acquires the sets of robots selected to move upward and downward, respectively. At lines~\ref{l:computeTargets1}-\ref{l:computeTargets2}, a robot computes the targets of all the robots. At line ~\ref{l:checkConflicts}, the robot checks if it is not selected to move upward and if any couple of robots have the same target. This test avoids collisions. Possible conflicting moves are shown in Figure~\ref{fig:movements}.(b). Two robots can have the same target when they are in the same column at distance two and the robot with the smallest row index is selected to move downward, while the other upward. An example is shown in Figure~\ref{fig:movements}.(b) for robots $r_3$ and $r_4$. The only other possible collision is for the robot $r_1$ having $t(r_1)=(2,2)$ (case five in Equation~(\ref{formula})). There might be a robot $r_2$ with $l(r_2)=3$ and $c(r_2)=2$ selected to move upward. This configuration is shown in Figure~\ref{fig:movements}.(b). In all these cases, to avoid any collision, the upward movement is performed only when there are no robots having the same target, otherwise the robot stays still. Each conflict is resolved in a robot cycle since downward and side movements are always allowed. 

Figure~\ref{fig:movements} shows the three types of possible movements performed by robots. Robots move concurrently without collisions. Figure~\ref{fig:movements}.(a) shows robots moving downward or upward and having different targets. Figure~\ref{fig:movements}.(b) shows two robots having the same target. To resolve the conflict,
the upward movement is stopped for a cycle. 
Figure~\ref{fig:movements}.(c) shows the cases in which a robot is selected to move upward ($r_8$) or downward ($r_5$) on a target vertex that is already occupied by another robot ($r_7$, $r_6$ respectively). Robots $r_5$ and $r_8$ perform their move while $r_6$ and $r_7$ move on a neighbouring vertex on the same row and closer to the center of $\GSmn$. 
Since movements are concurrent (robots are
synchronous), collisions are avoided. 

\block{Task $T_3$} This task is designed to bring $n-1$ robots to their final target except for $r_g$. This task activates when task $T_2$ is over, therefore $\pre_3$ holds:
$$\small \pre_3  \equiv \RV=1 \wedge \forall \textrm{ row } l: (\Bl=0 \wedge \Al=0) $$
Given the embedding of $F$ on $\GSmn$, in each row $l$, there are $t_l$ targets and $n_l$ robots, with $t_l$=$n_l$, therefore robots identify their final target and move toward it without collisions.
Given the particular shape of $F$, there are at most two targets per row, therefore we can state the move $m_3$ as follows: 
\textit{for each row, the rightmost robot moves toward the rightmost target and the leftmost robot moves toward the leftmost target except for $r_g$.}

\block{Task $T_4$} During task $T_4$, the guard $r_g$ moves from the corner of $\GSmn$ and goes toward its final target. This task activates when $\pre_4$ holds: 
$$\small \pre_4  \equiv n-1 \textrm{ robots but } r_g \textrm{ match their final target}.$$

The corresponding move is called $m_4$ and is defined as follows: \textit{$r_g$ moves toward its final target}. The embedding of $F$ guarantees that the final target of $r_g$ is on its neighbouring vertex on row 1. Therefore, in one step, $r_g$ reaches its target.    
After task $T_4$, the pattern is completed.

\block{Task $T_5$} This is the task in which each robot recognizes that the pattern is formed and no more movements are required. Each robot performs the null movement keeping the current position. The precondition is
$$\small \pre_5  \equiv F \textrm{ is formed}.$$

Although our algorithm is designed so as to form a specific pattern $F$ that solves $\GMVopt$, $\pre_5$ could be simply stated as `$\GMVopt$ solved'. In this way, robots would stop moving as soon as the problem is solved and not necessarily when the provided pattern $F$ is formed. However, since the formation of $F$ is usually required, for the ease of the discussion we prefer the current form for $\pre_5$.


\begin{table}[t]
\centering
\small
\begin{tabular}{|l|l|l|c|}
\hline
\textit{sub-problems}                                                         & \textit{task} & \textit{precondition}                                                                                 & \textit{transitions} \\ \hline
\multirow{3}{*}{Placement of the guard robot}                                 & $T_{1a}$      & true                                                                                                  & $T_{1a}$, $T_{1b}$       \\ \cline{2-4} 
                                                                              & $T_{1b}$      & $\RS \ge 1$ $\wedge$ $\RV=0$                                                                               & $T_{1b}$, $T_2$, $T_3$, $T_4$            \\ \cline{2-4} 
                                                                              & $T_{1c}$      & $\RV > 1$                                                                                             & $T_2$, $T_3$, $T_4$         \\ \hline
\begin{tabular}[c]{@{}l@{}}Bringing $t_l$ robots for each row\end{tabular} & $T_2$         & \begin{tabular}[c]{@{}l@{}}$\RV=1$ $\wedge$ $\exists$ $l \in \{1\ldots m\} : $\\ $\Bl\neq 0$ $\lor$ $\Al\neq 0$\end{tabular} & $T_2$, $T_3$, $T_4$   \\ \hline
Bring $n-1$ robots to final target                                                  & $T_3$         & \begin{tabular}[c]{@{}l@{}}$\RV=1$ $\wedge$ $\forall$ row $l$  ($\Bl=0$ $\wedge$ \\ $\Al=0$)  \end{tabular} & $T_3$, $T_4$         \\ \hline
Bring the guard robot to final target                                               & $T_4$         & $n-1$ robots on final target                                                                       & $T_5$         \\ \hline
Termination                                                                   & $T_5$         & $F$ formed                                                                                            & $T_5$         \\ \hline
\end{tabular}
\caption{\small The table summarizes the phases of the algorithm: the first column reports a summary of the task’s goal, the second column reports the task's name, the third column reports, for each task the precondition to enter the task, the last column reports the transitions among tasks.}
\label{tab:tasks}
\end{table}

\begin{figure}[t]
\begin{center}
\begin{tikzpicture}[thick, main/.style = {draw, circle}, node distance={26mm}] 

\node[main] (2) {$~T_{1a}$}; 
\node[main] (3) [right of=2] {$~T_{1b}$}; 
\node[main] (5) [above right of=3] {$~T_{2~}$}; 
\node[main] (7) [below right of=3] {$~T_{4~}$}; 
\node[main] (6) [above right of=7] {$~T_{3~}$}; 
\node[main] (8) [right of=7] {$~T_{5~}$}; 
\node[main] (4) [right of=6] {$~T_{1c}$}; 

\draw[-latex] (2) to [out=120,in=90,looseness=4] (2);
\draw[-latex] (2) -- node[midway, above right, sloped, pos=0] {} (3);
\draw[-latex] (3) to [out=120,in=90,looseness=4] (3);
\draw[-latex] (3) -- node[midway, above right, sloped, pos=0] {} (5);
\draw[-latex] (3) -- node[midway, above right, sloped, pos=0] {} (6);
\draw[-latex] (3) -- node[midway, above right, sloped, pos=0] {} (7);
\draw[-latex] (4) -- node[midway, above right, sloped, pos=0] {} (5);
\draw[-latex] (4) -- node[midway, above right, sloped, pos=0] {} (6);
\draw[-latex] (4) -- node[midway, above right, sloped, pos=0] {} (7);
\draw[-latex] (5) to [out=120,in=90,looseness=4] (5);
\draw[-latex] (5) -- node[midway, above right, sloped, pos=0] {} (6);
\draw[-latex] (5) -- node[midway, above right, sloped, pos=0] {} (7);
\draw[-latex] (6) to [out=120,in=90,looseness=4] (6);
\draw[-latex] (6) -- node[midway, above right, sloped, pos=0] {} (7);
\draw[-latex] (7) -- node[midway, above right, sloped, pos=0] {} (8);

\draw[-latex] (8) to [out=120,in=90,looseness=4] (8);

\end{tikzpicture}
\end{center}
\caption{Transition graph (derived from Table \ref{tab:tasks}).}
\label{fig:transitions}
\end{figure}
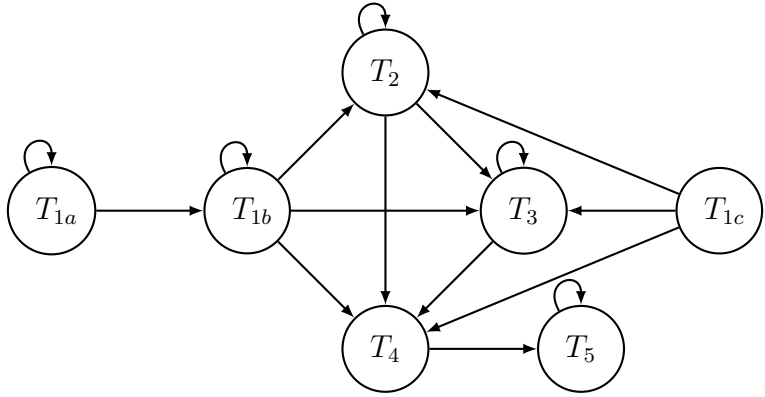

\subsection{Formalization and correctness}\label{sec:correctness-grid}
We have already remarked that the algorithm has been designed according to the methodology recalled in Section~\ref{sec:methodology}. Accordingly, Table~\ref{tab:tasks} summarizes the designed tasks, the corresponding preconditions, and the possible transitions from each task. Furthermore, all the transitions are shown in the transition graph depicted in Figure~\ref{fig:transitions}. 

We observe that the predicates used in the algorithm are all well-formed since they guarantee that $\Prop_1$, $\Prop_2$, and $\Prop_3$ are all valid. In particular, $\Prop_1$ follows from the definition of the simple preconditions expressed in Table~\ref{tab:tasks}, $\Prop_2$ holds because each predicate $P_i$ has been defined as indicated in Equation~\ref{eq:predicates}, and $\Prop_3$ directly follows from the definitions of $P_i$ (if $P_5, P_4, \ldots, P_{1b}$ are all false, then $P_{1a}$ holds). 
 
Concerning the correctness of $\algoAsym$, still using the methodology in the remainder of this section we show that properties $\h_1$, $\h_2$, and $\h_3$ hold by providing a specific lemma for each task. Finally, such lemmata will be used in a final theorem responsible for assessing the correctness of $\algoAsym$.

\begin{lemma}\label{lem:corr-T1a}
Let $C$ be a  configuration in $T_{1a}$. From $C$, $\algoAsym$ eventually leads to a configuration belonging to $T_{1b}$.
\end{lemma}

\begin{proof}
In task $T_{1a}$, Algorithm $\algoAsym$ selects a robot denoted as $r_g$, called guard, such that $D(r)$ is maximum and with the minimum view in case of ties.
Let us analyze properties $\h_i$, for $1\le i\le 3$, separately.

\begin{description}
\item[$\h_1$:] 
In task $T_{1a}$, no robots are on a side of the grid $\GSmn$, nor on its corners and $r_g$ moves toward the closest side of $\GSmn$, $D(R)$ increases for $r_g$, therefore, $r_g$ is repeatedly selected. When $r_g$ reaches a side of $\GSmn$, it is the only robot on a side of $\GSmn$ and $\RS=1$. Still, there are no robots on corners of $\GSmn$ therefore $\RV=0$, $\pre_{1b}$ becomes $\true$ and the configuration is in $T_{1b}$, since the preconditions of all the other tasks, except for $T_5$, require at least one robot on a corner of $\GSmn$, and $T_5$ requires more than one robot on the sides of $\GSmn$.

\item[$\h_2$:]
At each cycle, $r_g$ decreases its distance from the closest side of $\mbr(C)$ by one. Therefore, within a finite number of $\LCM$ cycles, it reaches its target and the configuration is not in $T_{1a}$ anymore.  

\item[$\h_3$:] 
Since $r_g$ is the robot such that $D(r)$ is maximum, it must be on a side of $\mbr(C)$. While moving toward the closest side of $\GSmn$, $r_g$ increases its distance from the other robots therefore it cannot meet any other robot on its way toward the target and no collision can occur.
\end{description}
\end{proof}

\begin{lemma}\label{lem:corr-T1b}
Let $C$ be a  configuration in $T_{1b}$. From $C$, $\algoAsym$ eventually leads to a configuration belonging to $T_2$, $T_3$ or $T_{4}$.
\end{lemma}

\begin{proof}
In task $T_{1b}$, Algorithm $\algoAsym$ selects a robot denoted as $r_g$ on the perimeter of $\GSmn$, closest to a corner of $\GSmn$, and having the minimum view in case of ties. Let us analyze properties $\h_i$, for $1\le i\le 3$, separately.

\begin{description}
\item[$\h_1$:] 

At the beginning of the task, there are no robots on a corner of $\GSmn$, and $r_g$ moves toward the closest corner. As $r_g$ moves toward its target, the distance from it decreases, therefore $r_g$ is repeatedly selected. When it reaches its target, there is a single robot on a corner of $\GSmn$ and $\RV=1$. Then, the obtained configuration can be in $T_2$, $T_3$ or $T_4$, all configurations in which the $r_g$  is placed on a corner of $\GSmn$. The obtained configuration is not in $T_5$ because the pattern $F$ has no targets on the corners of $\GSmn$.

\item[$\h_2$:]
At each cycle, $r_g$ decreases its distance from the closest corner of $\GSmn$ by one. Therefore, within a finite number of $\LCM$ cycles, $r_g$ reaches its target and the configuration is not in $T_{1b}$ anymore.

\item[$\h_3$:] 

Since $r_g$ is the robot closest to the corner of $\GSmn$ it cannot meet any other robot on its way toward the target and no collision can occur.
\end{description}
\end{proof}

\begin{lemma}\label{lem:corr-T1c}
Let $C$ be a  configuration in $T_{1c}$. From $C$, $\algoAsym$ eventually leads to a configuration belonging to $T_2$, $T_3$ or $T_{4}$.
\end{lemma}

\begin{proof}
In task $T_{1c}$, Algorithm~\ref{alg:serpentone} moves robots along \specialpaths. Let $p$ be the number of fully-occupied \specialpaths. $p$ cannot be greater than two and it can be two only when $k=\min(N,M)=4$. In fact, the length of a \specialpath is $k^2/4$ when  $k$ is even and $(k^2-1)/4$ when $k$ is odd, whereas the maximum number of robots is $2k$. For $k$ even, we have that 
$pk^2/4=2k$, that is $pk=8$. Hence, if $k>4$ there can be only one fully-occupied \specialpath, otherwise $k=4$ and there can be two fully-occupied \specialpaths. Similar analysis can be done for $k$ odd that leads to $pk<8$, then there can be only one fully-occupied \specialpath.

When $p=2$, the \specialpaths must be on successive corners of $\GSmn$ otherwise the configuration would be symmetric.  
Let us analyze properties $\h_i$, for $1\le i\le 3$, separately.
\begin{description}
\item[$\h_1$:] 
When $T_{1c}$ starts, $\RV \ge 1$. After the move, the guard $r_g$ is placed and $\RV=1$. Therefore, the configuration is either in $T_2$, $T_3$ or $T_4$ and it is not in $T_5$ because the pattern $F$ has no targets on corners of $\GSmn$.

\item[$\h_2$:]

In task $T_{1c}$, all the corners of $\GSmn$ but one are emptied in a robot cycle. 

\item[$\h_3$:] 

The \specialpaths are designed so that they are disjoint. During task $T_{1c}$, only robots on \specialsubpaths move along the \specialpath. These are the robots on a corner of $\GSmn$ and the ones in front of it until the first empty vertex. Since robots are 
synchronous, all these robots move forward by an edge, hence no collision can occur.
\end{description}
\end{proof}

\begin{lemma}\label{lem:corr-T2}
Let $C$ be a configuration in $T_{2}$. From $C$, $\algoAsym$ eventually leads to a configuration belonging to $T_3$ or $T_{4}$.
\end{lemma}

\begin{proof}
During task $T_2$, robots move to place two robots per row. The only exception occurs when $n$ is odd, in which case the last row requires just one robot. In particular, each robot runs Algorithm~\ref{alg:move} in which they recall Algorithm~\ref{alg:selection} that selects the robots moving upward and downward for each row. The first row is identified by the position of $r_g$ on the upper-right corner of $\GSmn$. Let us analyze properties $\h_i$, for $1\le i\le 3$, separately.
\begin{description}
\item[$\h_1$:]

The choice of robots and their movements avoid robots occupying more than a corner of $\GSmn$. 
Indeed, Algorithm~\ref{alg:selection} selects robots moving upward and downward.
When the number of rows $M$ of the grid $\GSmn$ are equal to $\lceil n/2 \rceil$, the algorithm selects robots between the second and the $(N-1)$-th column.
The number of robots on the grid ensures that, even in a configuration in which robots in each row, from the second to the $(M-1)$-th one, occupy the first and the last columns, there are at least other two robots if $n$ is odd and three if $n$ is even that can be selected to move, able to finalize task $T_2$. Since no robots can move on a corner of $\GSmn$, then the configuration is not in $T_{1a}$, $T_{1b}$ nor in $T_{1c}$.

When $M>\lceil n/2 \rceil$, robots do not move toward the last row of $\GSmn$, therefore they cannot occupy the corners of the $M$-th row of $\GSmn$. 

If a robot $r_1$ occupies the vertex with coordinates $(2,1)$, $RU_2=1$, and it is the only robot on row 2,
to avoid occupying the corner of $\GSmn$ with coordinates $(1,1)$, the target of $r_1$ is $(2,2)$ according to the fifth case of Equation~(\ref{formula}). 
If a robot $r_2$ occupies the vertex with coordinates $(2,N)$  and it is selected to move upward, $r_2$ moves on its target $(1,N)$ while the guard robot $r_g$ moves to coordinates $(1,N-1)$ according to the third case of Equation~(\ref{formula}). 
Then, the role of $r_g$ is taken by robot $r_2$ and a single corner of $\GSmn$ is occupied by a robot.  
In both cases, the configuration is not in $T_{1_c}$ because a single corner of $\GSmn$ remains occupied. Moreover, the configuration is neither in $T_{1a}$ nor $T_{1b}$ since $r_g$ is not moved, except for the case in which it is replaced by another robot.

During task $T_2$, the guard $r_g$ is placed on a corner of $\GSmn$  and $\RV=1$. At each cycle, $B_l$ becomes 0 for the first row $l$ for which $B_l \neq 0$. In at most $M-1$ steps, $B_l=0$ and $A_l=0$ for each $l$ in at most $2(M-1)$ cycles, given that the upward movement can be prevented for a cycle when two robots have the same target.
Examples of robots having the same target are depicted in Figure~\ref{fig:movements}.(b). In both cases, the algorithm stops any upward movement, while allowing side and downward movements, see line~\ref{l:checkConflicts} of Algorithm~\ref{alg:move}.  At the successive cycle, robots are on the same column and both move.
Once solved, no other conflict can occur in the same row. Then, in a finite number of cycles, $A_l$ becomes 0 for each $l$. At the end of the task, there are two robots on each row except when $n$ is odd, in which case the last row contains a single robot.
Precondition $\pre_3$ becomes $\true$, eventually, and the configuration is in $T_3$. If $n-1$ robots match their target, the configuration is in $T_4$ and it is not in $T_5$ because the pattern $F$ has no targets on corners of $\GSmn$.

\item[$\h_2$:]

As described in $H_1$, at the end of this task, $\Bl=0$ $\wedge$ $\Al=0$ for each row $l$. This condition is reached  in at most $2(M-1)$ cycles since the upward and downward movements are concurrent and no other configuration will be in $T_2$ anymore. 

\item[$\h_3$:] 

When the number of robots selected to move downward on row $l$ is such that $RD_l\ge 2$, the exceeding number of robots on $l$ will saturate all the targets of row $l+1$. Therefore, in the same cycle, any robots on row $l+1$ occupying the targets of robots on row $l$ must also move downward. As a consequence, any robot selected to move downward on row $l$ will reach a free target.
When $RD_l=1$, a robot $r$ moves on row $l+1$ and at the same time, the robots on row $l+1$ will also move downward leaving at most one robot $r_1$. If $r$ is on the same column of robot $r_1$, $r$ moves downward while $r_1$ moves to its neighbour closer to the center of $\GSmn$ (see Figure~\ref{fig:movements}.(c)). Note that, the neighbours of $r_1$ will be empty since all other robots on row $l+1$ left the row.
Moreover, the choice of the neighbour toward the center avoids $r_1$ going to one of the corners of $\GSmn$, see  cases three and four of Equation~(\ref{formula}). 
The same reasoning applies to robots moving upward.
When there are robots having the same target, see robots in Figure\ref{fig:movements}.(b) for reference, the algorithm detects this condition at line~\ref{l:checkConflicts}, and the upward movement is not performed. The robots are allowed to move downward or to the side, therefore collisions are avoided.
\end{description}
\end{proof}

\begin{lemma}\label{lem:corr-T3}
Let $C$ be a  configuration in $T_3$. From $C$, $\algoAsym$ eventually leads to a configuration belonging to $T_4$.
\end{lemma}

\begin{proof}
Task $T_3$ is designed to bring $n-1$ robots to their final target on $F$ except for $r_g$. 
Let us analyze properties $\h_i$, for $1\le i\le 3$, separately.
\begin{description}
\item[$\h_1$:] 

During this task, there are $r_l$ robots and $t_l$ targets per row. In each row, robots move toward their final target on their same row. The task continues until $n-1$ robots are correctly placed according to the pattern, $\pre_4$ becomes true and the configuration is in $T_4$.

\item[$\h_2$:]
At each $\LCM$ cycle, in each row, robots reduce the distance from their target by one until they reach the target. 

\item[$\h_3$:] 
There are at most two robots per row and two targets per row. Therefore, the rightmost robot goes to the rightmost target and the leftmost robot goes toward the leftmost target. In this way, collisions are avoided.
\end{description}
\end{proof}

\begin{lemma}\label{lem:corr-T4}
Let $C$ be a  configuration in $T_4$. From $C$, $\algoAsym$ eventually leads to a configuration belonging to $T_5$.
\end{lemma}

\begin{proof}
In task $T_4$, $n-1$ robots are correctly positioned according to the pattern except for $r_g$. From this configuration, $r_g$ moves toward its final target, in a single $\LCM$ cycle. Let us analyze properties $\h_i$, for $1\le i\le 3$, separately.
\begin{description}
\item[$\h_1$:] 

As $r_g$ moves, it matches its target on $F$, then the pattern is formed, $\pre_5$ becomes $\true$ and the configuration is in $T_5$.

\item[$\h_2$:]
The embedding of the pattern $F$ guarantees that the target of $r_g$ is at distance one from the corner of $\GSmn$ in which it resides, therefore in one \LCM cycle the task is over. 

\item[$\h_3$:] 
All robots, except for $r_g$, are matched and perform the nil movement, no other robots are on the target of $r_g$ given the definition of $m_4$, therefore no collision can occur.
\end{description}
\end{proof}

In the following, we state our main result in terms of time required by the algorithm to solve the problem $\GMVopt$.
Time is calculated using the number of required \LCM cycles 
given that robots are synchronous. Let $L$ be the side of the smallest square that can contain both the initial configuration and target configuration. 
Note that, any algorithm requires at least $O(L)$ \LCM cycles to solve $\GMVopt$.
Our algorithm solves $\GMVopt$ in $O(L)$ \LCM cycles which is time optimal. Our result is stated in the following theorem:

\begin{theorem}\label{teo:iff}
$\algoAsym$ is a time-optimal algorithm that solves $\GMVopt$ in each asymmetric configuration $C$ defined on a finite grid. 
\end{theorem}

\begin{proof}
Lemmata~\ref{lem:corr-T1a}-\ref{lem:corr-T4} ensure that properties $\h_1$, $\h_2$, and $\h_3$ hold for each task $T_{1a}$, $T_{1b}, \ldots, T_5$. Then, all the transitions are those reported in Table~\ref{tab:tasks} and depicted in Figure~\ref{fig:transitions}; the generated configurations can remain in the same task only for a finite number of cycles; and the movements of the robots are all collision-free. Lemmata~\ref{lem:corr-T1a}-\ref{lem:corr-T4} also show that from a given task only subsequent tasks can be reached, or $\pre_5$ eventually holds (and hence $\algoAsym$ is solved). 
This formally implies that, for each initial configuration $C$ and for each execution $\Ex : C=C(t_0),C(t_1),C(t_2),\ldots$ of $\algoAsym$, there exists a finite time $t_j>0$ such that $C(t_j)$ is similar to the  pattern to be formed in the $\GMVopt$ problem and $C(t_k) = C(t_j)$ for each time $t_k\ge t_j$. 

Concerning the time required by $\algoAsym$, it is calculated using the number of required \LCM cycles, as robots are synchronous. Recall that $L$ is the side of the smallest square that can contain both the initial configuration and the target configuration. 
Tasks $T_{1a}$ and $T_{1b}$ require $O(L)$ \LCM cycles since a robot must move for $O(\max\{N,M\})$ edges in each of them. Task $T_{1c}$ requires exactly one \LCM cycle. By the proof given in Lemma~\ref{lem:corr-T2}, robots complete Task $T_2$ in at most $2(M-1)$ \LCM cycles, that is in $O(L)$ time.
Task $T_3$ requires at most $O(N)$ \LCM cycles, i.e. $O(L)$ time. Task $T_4$ requires exactly one \LCM cycle. Then, Algorithm $\algoAsym$ requires a total of $O(L)$ \LCM cycles, hence it is time optimal since no algorithm can solve $\GMVopt$ in less than $O(L)$ \LCM cycles.

\end{proof}

\subsection{The case of symmetric configurations and infinite grids}\label{sec:infinite-grids}
In this section, we discuss (1) how $\algoAsym$ can be extended to a general algorithm $\algo$ able to handle also symmetric configurations, and (2) how, in turn, $\algo$ can be modified into an algorithm $\algoinfty$ that solves the same problem defined on the infinite grid $\GS$. 

\block{Symmetric configurations}
We first explain how to solve symmetric initial configurations with $\rho(C)=1$, then those with  $\rho(C)\in \{2,4\}$. If $C$ is a symmetric configuration with $\rho(C)=1$, then there exists a robot $r_c$ located at the center $c$ of $C$, and for $C'=\{C \setminus r_c\}$, $\rho(C')\in \{2,4\}$. To make the configuration asymmetric, $\algo$ must move $r_c$ out of $c$ (symmetry-breaking move). To this end, when $r_c$ has an empty neighbour -- arbitrarily chosen if more than one -- then $r_c$ moves toward it. If all the four neighbours of $r_c$ are occupied but there is at least an empty vertex on the same row or column of $r_c$, the neighbors of $r_c$ and the robots in front of them until the first empty vertex, move along the row or column. As a result, a neighbour of $r_c$ will eventually be emptied. Then, the symmetry-breaking move can be applied. If all the vertices on the same row and column of $r_c$ are occupied, then all other vertices except one (if any) must be empty. Therefore the four neighbor robots of $r_c$ move toward a vertex placed on the right with respect to $c$, if empty. Again, a neighbour of $r_c$ will eventually be emptied and the symmetry-breaking move can be applied. When the configuration is made asymmetric, $\algoAsym$ runs on $C$ and $\GMVopt$ is solved.

Consider now $C$ with $\rho(C)\in \{2,4\}$. In these cases, the configurations is divided into rectangular sectors, i.e., regions of $\GSmn$ which are equivalent to rotations. Then, $\algo$ instantiates $\algoAsym$ in each sector according to suitably chosen patterns. 

We now explain how the subdivision into sectors is performed. Given the symmetry of the configuration, the algorithm $\algo$ selects $\rho(C)$ robots as guards and places each of them on different corners of the grid. The placement is done as in $\algoAsym$ by means of either tasks $T_{1a}$ and $T_{1b}$ or $T_{1c}$. Given the placement of the guards, robots identify and enumerate rows and columns, as done in Section~\ref{alg-highlevel},
and agree on how to subdivide $\GSmn$ into $\rho(F)$ sectors according to values of $\rho(C)$, of $|R|\mod4$, and the type of center $tc(C)$. 

\begin{figure}[t]
     \centering
     \includegraphics[scale=0.5]{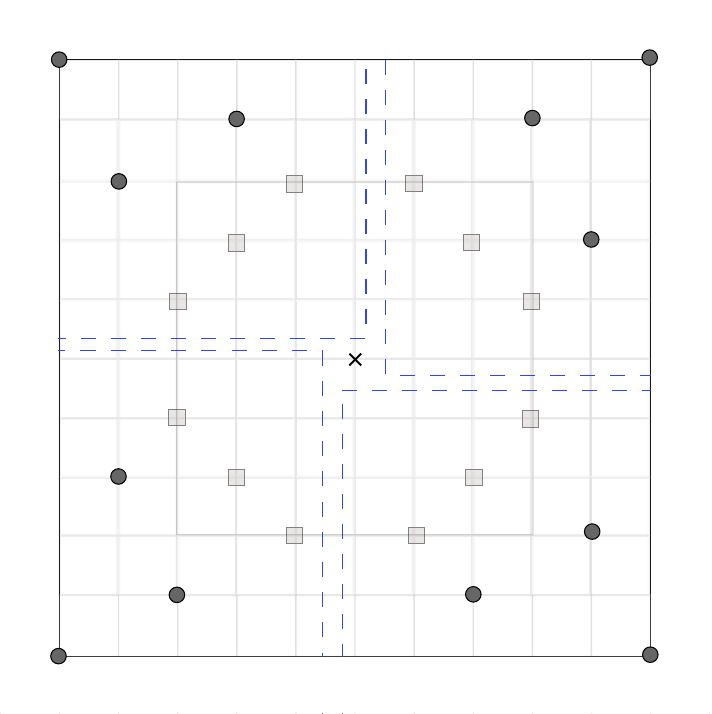}
     \includegraphics[scale=0.5]{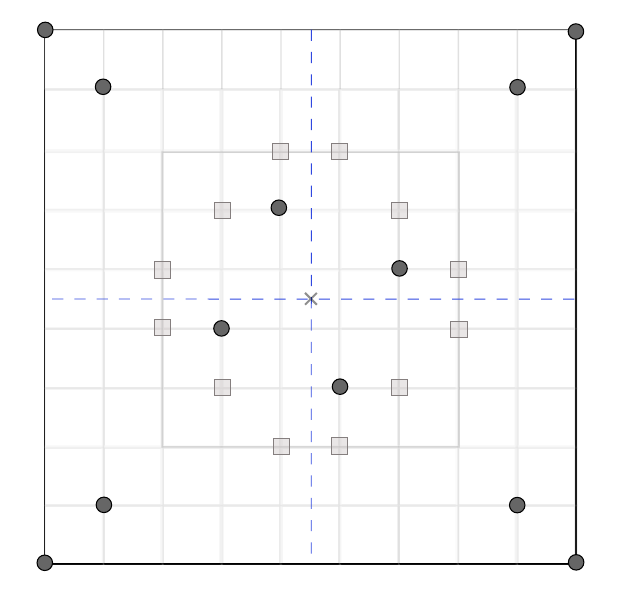}~~~~~~~
         
        \caption{\small \centering Patterns $F$ for $\rho(C)=4$: left $\tc(C)=1$, right $\tc(C)=3$.
}
         \label{fig:confs-rho4}
\end{figure}

\begin{figure}[h]
     \centering
         \includegraphics[scale=0.5]{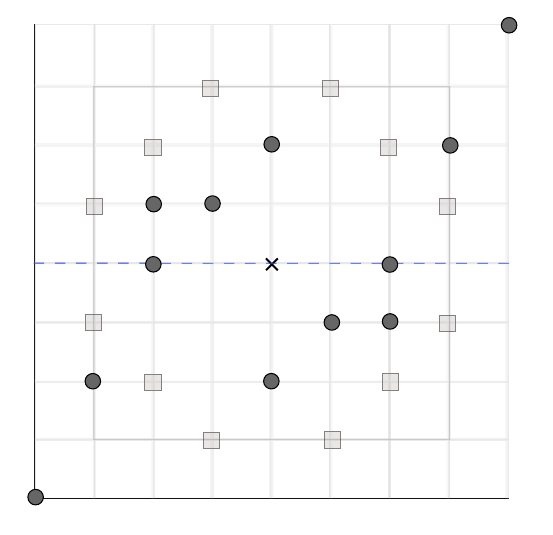}
         \includegraphics[scale=0.5]{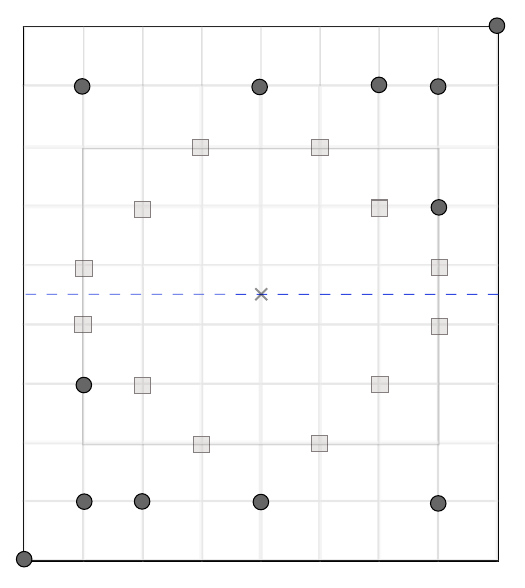}
         \includegraphics[scale=0.5]{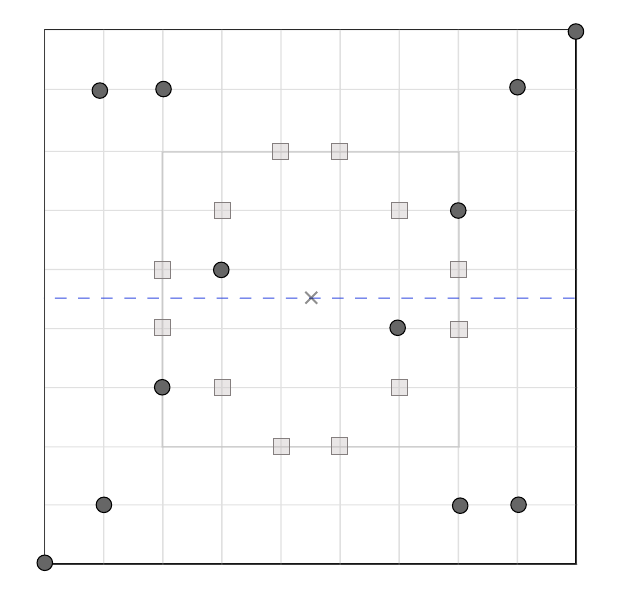}
        \caption{\small \centering Patterns $F$ for $\rho(C)=2$ and $|R|\mod 4 = 0$: left $\tc(C)=1$, middle $\tc(C)=2$, right $\tc(C)=3$.}
         \label{fig:confs-rho2-resto-0}
\end{figure}

\begin{figure}[h]
     \centering
         \includegraphics[scale=0.5]{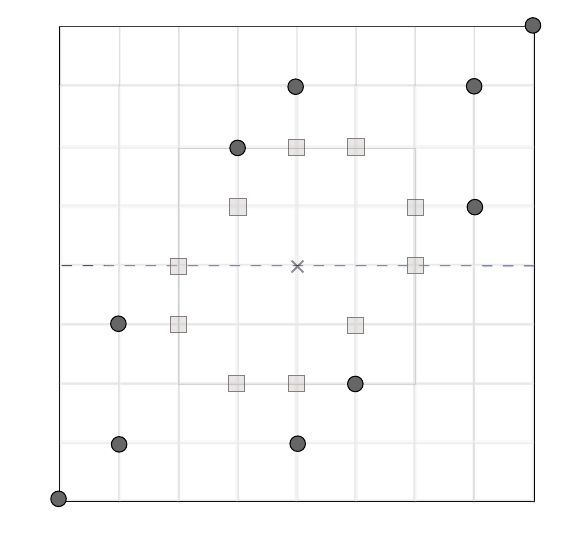}
         \includegraphics[scale=0.5]{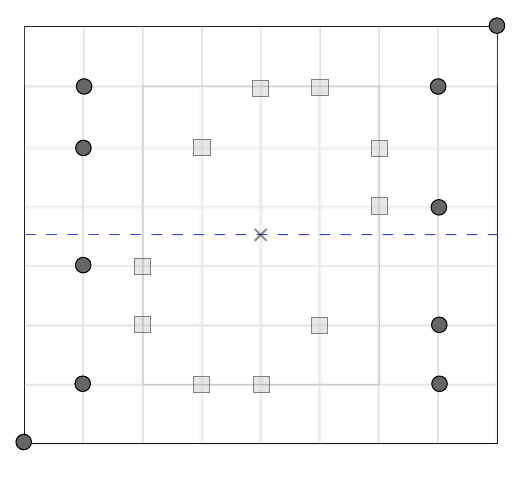}
         \includegraphics[scale=0.5]{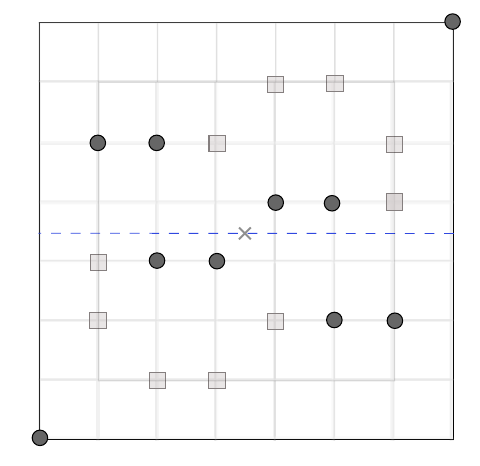}
        \caption{\small \centering Patterns $F$ for $\rho(C)=2$ and $|R|\mod 4 = 2$: left $\tc(C)=1$, middle $\tc(C)=2$, right $\tc(C)=3$.}
         \label{fig:confs-rho2-resto-2}
\end{figure}

For configurations having $\rho(C)=4$ the configuration is divided into four disjoint sectors (cf. Figure~\ref{fig:confs-rho4}): for centers $tc(C)=1$, each orthogonal line originating from the center is associated to the sector on its left, for centers of type $tc(C)=3$, sectors are obtained with two orthogonal lines passing through the center of the configuration. 
When $\rho(C)=2$ two sectors are obtained with a line parallel to the rows of $\GSmn$ passing through $c$ (cf. Figures~\ref{fig:confs-rho2-resto-0} and~\ref{fig:confs-rho2-resto-2}). Note that, when $tc(C)=1$, the line belongs to both sectors. 
In so doing, sectors keep a rectangular shape and $\algoAsym$ can be applied to each of them.

\medskip
We now explain how patterns are selected and embedded. 
Figures~\ref{fig:confs-rho4},~\ref{fig:confs-rho2-resto-0}, and~\ref{fig:confs-rho2-resto-2} also illustrate some examples concerning the optimal patterns for all cases and for specific values of $n$. 
From those examples, patterns $F$ for larger values of $n$ can be easily obtained by suitably enlarging the provided patterns (detailed instructions can be found 
in~\cite{D22}). 
Since robots in $C$ are synchronous, irrespective of the algorithm operating on $C$, the center $c$ of the configuration is invariant, therefore robots agree on the embedding of $F$ by identifying its center with $c$ and placing the $\rho(F)$ corners of $F$ with the maximum view closest to the $\rho(C)$ guard robots. $F$ is selected so that $\rho(C)$ divides $\rho(F)$ and $\tc(F)=\tc(C)$, and the placement of robots in $F$ solves $\GMVopt$. 

As pointed out before, each sector contains a sub-configuration that is asymmetric, then $\algo$ instantiates $\algoAsym$ in each sector while the definitions of functions $A_l, B_l, RD_l$, and $RU_l$ apply to each sector.  
Note that the algorithm works correctly even for configurations having $\rho(C)=2$ and $\tc(C)=1$ where the two sub-grids, in which the $\algoAsym$ runs independently, share the central row of $\GSmn$. 
In particular, the number of robots and targets is computed by each instance of $\algoAsym$ only considering those lying on the half of the central row closest to the guard. Note that, the center is never considered by the computation as there is neither a target nor a single robot there. If a robot from a sector, say the first one, moves on the central row, it may fall into the half row belonging to the other sector, say the second one, but its equivalent robot in the second sector would move in the opposite direction entering the half row belonging to the first sector and the number of robots in each sector is kept. In such situations, two robots may move and collide on the center of $C$, $c(R)$. In this case, we need to slightly modify $\algoAsym$: robots are prevented from moving on $c(R)$ while other robots will eventually move on the central row. 

Another difference with $\algoAsym$, is the movement of the guard robot toward its final target in $F$ during task $T_4$. In this case, $r_g$ is not one step away from its target as in the asymmetric case, given the embedding of $F$ into the center of $\GSmn$. Move $m_4$ works also in $\algo$, but it is completed in more than one \LCM cycle.

\block{Infinite grids}
To obtain $\algoinfty$, it is sufficient to make small changes to tasks $T_{1a}$, $T_{1b}$, and $T_4$. In $\algoAsym$, task $T_{1a}$ selects a single robot $r_g$ to occupy a corner of $G$. Since $\GS$ does not have corners, $\algoinfty$ selects $r_g$ as in $T_{1a}$ and then moves it to a distance $\mathcal{D}\ge 3\cdot \max\{w(C'),w(F)\}$, where $C'=\{C\setminus r_g\}$, and $w(C')$, $w(F)$ are the longest sides of $\mbr(C')$ and $\mbr(F)$, respectively. In task $T_{1b}$, $r_g$ must be chosen as the robot with a distance $\mathcal{D}$ from $C'$, and it moves toward a corner of $C$. In $T_2$, the first row is identified as the first row of $C'$ occupied by a robot, approaching $C'$ from $r_g$. The embedding on $F$ is achieved by matching the corner of $F$ with the maximum view in correspondence with the corner of $C'$ on the first row and having the same column of $r_g$. Tasks $T_2$ and $T_3$ are unchanged, while in task $T_4$, $r_g$ takes $\mathcal{D}$ \LCM cycles to move toward its final target in $F$.

\section{Conclusion}
\label{sec:concl}

We have studied the \textsc{Geodesic Mutual Visibility} problem in the context of robots moving along the edges of a (finite or infinite) grid and operating under the \LCM model. Regarding capabilities, robots are rather weak, as they are oblivious and without any direct means of communication. Robots are considered to be synchronous and endowed with chirality. We have shown that $\GMVopt$ can be solved by a time-optimal distributed algorithm.

This work opens a wide research area concerning $\GMV$ on other graph topologies or even on general graphs. However, difficulties may arise in moving robots in the presence of symmetries. Then, the study of $\GMV$ in asymmetric graphs or graphs with a limited number of symmetries deserves main attention.
Other directions concern deeper investigations into the different types of schedulers: synchronous, semi-synchronous or asynchronous.


\end{document}